\documentclass{kapproc} 
\kluwerbib %

\begin{document}

\articletitle{Bare strange quark stars:\\ Formation and emission}

\author{Renxin Xu}
\affil{School of Physics, Peking University, Beijing 100871,
       China; {\tt rxxu@bac.pku.edu.cn}}

\prologue{A theory is just a mathematical model, that describes
and codifies the observations.} {A man having a belief in {\sl
Positivist Philosophy}}

\begin{abstract}
Recent achievements of bare strange stars are briefly reviewed.
A nascent protostrange star should be bare because of strong mass
ejection and high temperature after the supernova detonation
flame, and a crust can also hardly form except for a
super-Eddington accretion.
The magnetosphere of a bare strange star is composed mainly of
$e^\pm$ pair plasma, where both inner and outer vacuum gaps work
for radio as well as high energy nonthermal emission.
A featureless thermal spectrum is expected since no ion is above
the quark surface, whilst electron cyclotron lines could appear in
some bare strange stars with suitable magnetic fields.
Various astrophysical implications of bare strange stars are
discussed.
\end{abstract}


\section{Introduction}

Identifying strange quark stars (or simply {\em strange stars},
SS) is among the most important problems in modern astrophysics.
Strange stars (SSs) are hypothetical compact objects that consists
of roughly equal numbers of deconfined up, down and strange
quarks, to affirm or negate the existence of which should have
profound implications in the study of the elemental strong
interaction (see, e.g., Xu 2002a, for a review).
It becomes more and more recognized in astrophysical community,
that those compact objects, previously thought as neutron stars
(NSs), may actually be SSs. Whatever the objects are, the most
essential and important thing is to find {\em clear} observational
signatures of SSs {\em or} NSs.

Bare strange stars (BSSs) are SSs with bare quark surfaces, i.e.,
without possible crusts of ordinary matter.
Recently, much attention is paid to the study of BSSs, because the
peculiarity of quark surface properties may eventually help us to
identify a quark star (e.g., Xu 2002a).
Although most of literatures are concerned with crusted strange
stars (e.g., Madsen 1999, for a review), several groups are doing
care about the study of the physics and astrophysics of BSSs.
Usov (2002, for a review) proposed a mechanism for $e^\pm$ (and
thus photon) emission in the superstrong electric field above the
surface of a hot BSS, and considered possible heating evens of
BSSs in order to explain some astronomical phenomena;
while Ray et al. (2000) calculated the mass-radius relations of
BSSs with finite temperatures (upto 70 MeV).
Due to the abrupt density change, from $\sim 4\times 10^{14}$
g/cm$^3$ to 0 in $\sim 1$ fm, BSSs may collapse into black holes
via microscopic black holes seeding (Gorham et al. 2002).
Motivated by the nonconservation of the quark axial current at MIT
bag boundary, Ng, Cheng \& Chu (2002) proposed a pion cloud, with
a thick of $\sim 1$ fm, around a BSS, which may affect
significantly the cooling curves of BSS.
However, if SSs appear as radio pulsars, plenty of data of which
are accumulated, what can the observation tell us about the nature
of pulsars?
It is necessary to combine the researches of radio pulsars, NSs,
and SSs (see, Xu, Zhang \& Qiao 2001a, for a review) all together
in order to have an effective study.

We will focus the review, in this paper, on the formation and
emission of BSSs, with the inclusion of various astrophysical
implications of BSSs.

\section{Formation of bare strange stars}

A protostrange star should be bare.
No numerical model of supernova explosion known hitherto has
included the conversion from protoneutron stars to protostrange
stars although such a conversion may be helpful in modelling the
burst process.
Nevertheless, some efforts have been made in trying to understand
the transition and combustion processes in detail.
From a kinetic point of view, Olinto (1987) for the first time
calculated the conversion of neutron stars into strange stars,
suggesting a deflagration mode with a burning velocity range from
$10^4$ km/s to a few cm/s. However it is found (Horvath \&
Benvenuto 1988; Benvenuto, Horvath \& Vucetich 1989; Benvenuto \&
Horvath 1989) that such slow modes are unstable. This instability
would be self-accelerated, and the burning should occur finally in
detonation modes although the transition from deflagration to
detonation has not been well understood (Lugones, Benvenuto \&
Vucetich 1994).
It is almost impossible that a crust can survive the detonation
flame.
First, the total released phase transition energy $\sim
10^{52}M_1$ ergs is much greater than the crust gravitational
binding energy $\sim 10^{48}M_1^2/R_6$ ergs, where $M_1$ is the SS
mass in $M_\odot$, $R_6$ the radius in 10 km.
Second, the rate of energy release when forming a strange star,
$\sim 10^{53}$ ergs/s, is much larger than the Eddington
luminosity, $\sim 10^{38}$ ergs/s.
In addition, high temperature ($\sim 30$ MeV) of newborn BSSs
increases the quantum penetration rate of ions (Usov 1998), and
the electric field can only thus sustain a crust with mass $\ll
10^{-5}M_\odot$.

Can a nascent BSS be covered via accretion in its future?
It was believed that, during and after supernova explosion,
material with ejection speed smaller than that of escape should
fall back to the center star with hypercritical accretion rate
(Chevalier 1989).
However whether this initial super-Eddington accretion occurs is
still a matter of debate because of the untractable nature of
modelling supernova explosion with the inclusion of rapid rotation
and strong magnetic field.
Owing to rapid rotation and strong magnetic fields, most of the
fallback matter may form temporarily a fossil disk, and the
accretion onto stellar surface can only be impossible when the
rotation period $P$ satisfies (Xu et al 2001a),
\begin{equation}
P > 3.2\times 10^3 B_{12}^{6/7} R_6^{18/7} M_1^{-5/7}\dot
M_{10}^{-3/7} ~{\rm s},
\end{equation}
where $B=10^{12}B_{12}$ G is the magnetic field, and $\dot
M=10^{10}\dot M_{10}$ g/s is the accretion rate. Also, a 3-D
magnetohydrodynamic simulations, not including stellar rotation
and B-fields, showed that the gravitational energy of the
infall-magnetized plasma has to be converted to other energies and
that the initial accretion rate migth be reduced significantly
(Igumenshchev \& Narayan 2002).
Furthermore, as argued by Xu (2002b), a crust can not form by
accretion as long as the accretion rate $<{\cal L}^*$,
\begin{equation}
{\cal L^*}\sim {9.1\times 10^{35}\over
\varepsilon}M_1R_6^{-1}P^{-1/2}~~{\rm ergs~s^{-1}},
\end{equation}
with $5\times 10^{-5}/P<\varepsilon<0.5$.
In conclusion, BSSs can exist in nature, some of which may be
active as radio pulsars, anomalous X-ray pulsars, soft
$\gamma-$ray repeators, etc. (Table 1).

\section{Magnetospheric and thermal emission}

BSS emission can be divided into magnetospheric and thermal parts.

One of the consequences of strong binding of charged particles
(quarks and electrons) on BSS surfaces is the formation of an
RS-type (Ruderman \& Sutherland 1975) vacuum inner gap above polar
cap (Xu \& Qiao 1998, Xu, Qiao \& Zhang 1999).
Although radio pulsar emission mechanism is not well understood,
the RS model is still the most popular one to connect
magnetospheric dynamics with general observations, with a ``user
friendly'' nature.
The RS model may naturally explain many observational features of
radio pulsars, such as drifting subpulses (e.g., Deshpande \&
Rankin 1999), microstructures (Hankins 1996), and even the
core-cone beams (Qiao \& Lin 1998) and their polarizations (Xu et
al. 2000a).
However, if radio pulsars are NSs with canonical dipole magnetic
fields, the RS model faces at least two difficulties: the binding
energy problem and the antipulsar issue.
These two can easily be overcome if BSSs are chosen as the nature
of radio pulsars.

Nonetheless the RS model may still work, but requires special
conditions, for neutron stars (Gil \& Mitra 2001, Gil, Melikidze
\& Mitra 2001, Gil \& Melikidze 2002).
Since ion cohesive energy is higher for a stronger magnetic field,
the binding energy crisis for NSs could be solved if a complicated
multipolar magnetic field, with a field strength being much higher
than that of the dipole component and a curvature radius being
much smaller than the stellar one, is assumed.
Whereas there are at least two points to be questioned in this
scenario: 1, the RS model can not work for antipulsars; 2, the
cohesive energy in reality could be much smaller than that of
condensed iron matter since the bombardment of backflow particles,
with an energy $>$ GeV in the frame of center of mass, may convert
heavy elements into lighter ones by nuclear reactions (Xu et al.
2000b).
Besides, the field configuration is also a mater of debate.

In addition, the existence of vacuum outer gap can also reflect
the strong binding of particles on BSS surface.
A plasma magnetosphere, with charge density of $\rho_{\rm GJ}$
(Goldreich \& Julian 1969) to quench $E_\parallel$ (the surface
electric fields being parallel to the magnetic fields) induced by
the unipolar effect, should surround a BSSs; any charge departure
from $\rho_{\rm GJ}$ has to result in acceleration, the region of
which is called ``gap''.
Owing to the force-free flow along spiral magnetic field lines
near light cylinder (Fig.4a in Holloway 1975) and the centrifugal
force, the charge density in the open-field-line region can {\em
not} be $\rho_{\rm GJ}$; gaps form then in this region.
Whereas the open-field-line region can in principal be divided
into two regions: region I with polar radius $r_{\rm p+}^{\rm d}$
and region II between field lines ``a'' and ``b'' (see Fig.1 of
Ruderman \& Sutherland 1975); the charge signs of these two are
opposite outside the light cylinder.
A natural way to close the electric current in BSS magnetospheres
is that an RS-type inner vacuum gap forms above the polar cap in
region I while an outer one (e.g., Cheng, Ho \& Ruderman 1986)
develops near the null surface in region II.
However there may be some unseemly points for the outer gap if
pulsars are not BSSs but NSs.
1. Because the field lines are equipotential, for a parallel
rotator as an example, electrons can freely flow from stellar
surface in all the open region; particles of the opposite sign may
emitted in a ``thin sheath'' (Michel 1975).
2. Since negatively charged particles flow out in region I,
positively charged ions in region II may flow freely from stellar
surface through the null surface if the binding energy of ions is
not high enough (Holloway 1975).
In both cases the outer vacuum gap can not work for NSs.

Actually observational supports (e.g., Romani \& Yadigaroglu 1995;
Wang, Xu \& Qiao 2002) could be hints of the existence of outer
gaps, and thus of the strong binding of charges on BSS surface.
Certainly it is very necessary to improve gap models of BSSs,
especially for the {\em interaction} between inner and outer
vacuum gaps; a {\em dynamically} consistent model with both kinds
of gaps may be essential to understand the observations.
It is worth noting that the null surface is not necessary to serve
as the inner boundary of the outer gap; it may be possible that
the inner boundary of the P-N junction (Holloway 1973) could be
much lower (Hirotani \& Shibata 2001).

Let's turn to the discussion of BSS thermal emission.
In principle, one can study the thermally radiative properties by
comparison of theoretically modelled spectra with that of
observations, in order to get a real information of photons from
quark matter astrophysically (whereas direct photons and lepton
pairs have been recognized to be the clearest signatures for
quark-gluon plasma in terrestrial physics, e.g., Cassing \&
Bratkovskaya 1999).
Unfortunately no emergent spectrum of BSSs appears in literature
although some efforts were tried (Chmaj et al. 1991, Page \& Usov
2002, Ng et al. 2002).
Nevertheless, as argued by Xu (2002b), the cooling and the thermal
radiation of a BSS may not be strongly conflict with observations
when polar cap heating is considered.
The total luminosity of BSS thermal emission, including photons
and $e^\pm$ pairs, was calculated (see Usov 2001a for the details)
in the frame of Usov (1998).

Xu (2002b) suggested that a featureless thermal spectrum could be
a probe for identifying strange stars, since no bound charged
particle is in discrete quantum states on the quark surface
without strong magnetic field; but discrete Landau levels appear
for charged particles in strong fields, which could result in a
cyclotron line spectrum.
In fact it is a central goal and a real competition among the
observers to find line emission in the thermal radiation of NS
atmospheres, since the stellar mass $M$ and radius $R$ may be
derived by obtaining its gravitational redshift (as $M/R$) and the
pressure broadening (as $M/R^2$) of the lines.
More advanced facilities, {\em Chandra} and {\em XMM-Newton}, make
this investigation possible.
Still almost no line is observed (Table 1) except for those
sources 1E 1207.4-5209, SGR 1806-20, and EXO 0748-676.
\begin{table}[ht]
\caption[]{``NSs'' with thermal X-ray spectrum observed by {\em
Chandra}, {\em XMM}, or others}
\begin{tabular*}{\textwidth}{@{\extracolsep{\fill}}lccc}
\sphline \it Name&\it Period {\rm (s)} &\it B-field {\rm (G)} &\it
Age {\rm (yr)} \cr \sphline
RX J1856.5-3754 (INS$^a$)&...&...&...\cr%
RX J0720.4-3125 (INS)&8.39&$\sim 10^{13}$&$\sim 10^6$\cr%
1E 1048.1-5937 (AXP$^b$) & 6.45 & Magnetar? &...\cr%
4U 0142+61 (AXP) & 8.69 & Magnetar? &...\cr%
PSR J0437-4715 (msPSR$^c$) & 0.00576 & $3\times 10^8$ & $4.9\times 10^6$?\cr%
PSR B0833-45 (Vela) & 0.0893 & $3.4\times 10^{12}$ & $1.1\times 10^4$\cr%
PSR B0656+14 &0.385&$4.7\times 10^{12}$&$1.0\times 10^5$\cr%
Kes 79 (CCO$^d$)&8.39&...&$0.6-1.2\times 10^4$?\cr%
Cas A (CCO)&0.0122?&$<5\times 10^{10}?$&...\cr%
Pup A (CCO, RXJ0822)&...&...&$\sim 4\times 10^3$\cr%
RX J1308.6+2127 (INS)&5.16&$1.7-3.2\times 10^{14}$?&...\cr%
RX J0806.4-4123 (INS)&11.37&$\sim 5\times 10^{10}$?&...\cr%
\sphline%
1E 1207.4-5209 (CCO, PKS1209)&0.424&$\sim 10^{11}?$&$(3-20)\times 10^3$?\cr%
SGR 1806-20 (SGR$^e$)&7.47&$\sim 5\times 10^{11}$?&...\cr%
EXO 0748-676 (X-ray burster)&...&...&...\cr%
\sphline%
\end{tabular*}
\begin{tablenotes}
References --- Zane et al. (2002), Seward et al. (2002), Murray et
al. (2001), Zavlin et al. (1999), Kaplan et al. (2002), Haberl \&
Zavlin (2002), Hambaryan et al. (2002), Sanwal et al. (2002), Xu
et al. (2003), Ibrahim et al. (2002), Cottam et al. (2002),
and those references related to Table 1 of Xu (2002b).\\
$^a$INS: isolated neutron stars. $^b$AXP: anomalous X-ray pulsars.
$^c$PSR: pulsars, msPSR: millisecond pulsars. $^d$CCO: compact
central objects in SNRs. $^e$SGR: soft $\gamma-$ray repeaters.
\end{tablenotes}
\end{table}

Sanwal et al. (2002) supposed a likely interpretation in which the
absorption features in 1E 1207.4-5209 are associated with atomic
transitions, and that it is very difficult to interpret the
absorption features in term of cyclotron lines.
However this opinion was criticized by Xu et al. (2003), who
proposed that 1E 1207.4-5209 may have a debris disk and is in a
propeller phase, with an accretion rate $\sim 6\times
10^{-11}M_\odot$/year.
Ibrahim et al. (2002) detected features in the bursts of a
soft-gamma-repeater SGR 1806-20, which is supposed as the proton
cyclotron lines in superstrong magnetic field ($\sim 10^{15}$ G).
Whilst, due to the high mass-energy ($\sim 1$ GeV) of a proton,
the ratio of the oscillator strength of the first harmonic to that
of fundamental in $10^{15}$ G is {\em only} $\sim 10^{-6}$! It is
not reasonable to detect the first and the {\em even} higher
harmonics. In fact, numerical spectrum simulations of atmospheres
with protons in superstrong fields have never show more than two
proton absorption lines (Ho \& Lai 2001).
Motivated by these, Xu et al. (2003) suggested that the observed
features could be electron cyclotron lines, and that SGR 1806-20
may have an ordinary magnetic field, $\sim 5\times 10^{11}$ G.
Also RX J0806.4-4123 shows a possible absorption at $0.4-0.5$ keV
(Haberl \& Zavlin 2002), which corresponds a B-field of $\sim
5\times 10^{10}$ G if being cyclotron-originated.
Why only these in which significant absorption features have been
detected so far? Xu et al. (2003)'s answer is that, for detectors
with observing energy from $\sim 0.1$ to $\sim 10$ keV, the
sensitive magnetic fields in which electrons can absorb resonantly
photons within that energy range are from $9\times 10^9$ G to
$1\times 10^{12}$G.
No objects with certain magnetic field strength, listed in the
upper part of Table 1, shows significate absorption feature.
It is worth noting that magnetospheric power law components of
BSSs are also featureless (Xu \& Qiao 1998, Xu et al. 2001a), but
a neutron star may have magnetospheric line features because of
the ions, pulled out from NS surface by the
space-charge-limited-flow mechanism, in the open field line
region.

Although BSSs may be common, SSs with their historical accretion
rates being much higher than the Eddington one could be crusted.
X-ray bursters, if being SSs, might be in this case. Cottam et al.
(2002) discovered significant absorption lines in the spectra of
EXO 0748-676, all with a readshift 0.35.
This result can not rule out an SS model for EXO 0748-676 (Xu
2002c).

\section{Conclusions and Discussions}

The formation of BSSs is reasonable, and the BSS thermal and
magnetospheric emission could be consistent with observations.
It is a hot topic now to discuss the astrophysical implications of
SSs (Dai \& Lu 1998, Cheng et al. 1998), whereas more attention
should be paid to the bare quark surface in the future.

There may be some other implications of BSSs.

1. Strong magnetic fields of radio pulsars (and others) are
essential to understand the observational data. This field could
be influenced significantly by, e.g., Hall drift and Ohmic decay
(Hollerbach \& R\"udiger 2002) in the crusts if pulsars are NSs or
SSs with crusts, but may keep constantly for BSS with color
superconductivity (Xu \& Busse 2001).

2. It is a strange and interesting thing that the masses of
observed stellar black hole candidates are $>\sim 7M_\odot$
whereas the mass limit of Fermion stars are $1-3M_\odot$. Why do
not we find black holes with $\sim 3-7M_\odot$?
This discrepancy may be explained if we assume that:
a, massive main-sequence stars have very rapid rotating cores in
the highly evolved phases;
b, quark matter in strange stars can be enough confined against
the centrifugal breaking.
In this case, it is found that the masses of stellar black holes
should be greater than $6M_\odot$ if a black hole can {only} form
when the stellar radius of a BSS is smaller than the radius of its
minimum stable orbit (Zhang, Zhang \& Xu 2003).

3. The soft $\gamma$-ray bursts of SGR 0526-66, with peak
luminosity $\sim 10^7L_{\rm Edd}$, needs ultra-strong field ($\sim
10^{17}$G) to constrain the fireball. An alternative binding is
through the quark surface; and it may be natural to explain the
bursting energy and the light curves in a framework that a
comet-like object falls to a BSS (Zhang et al. 2000, Usov 2001b).

4. It is suggested that the beam widths and polarizations of radio
pulsars can be used to derive their mass-radius relations (Kapoor
\& Shukre 2001), especially for the fastest rotating pulsar, PSR
1937+21 (Xu et al. 2001b), with the inclusion of general
relativistic effects.
Taking the simplest proposal that the inclination and impact
angles are $90^{\rm o}$ and $0^{\rm o}$, respectively, Xu et al.
(2001b) found stringent limits on the mass $M$ and the radius $R$:
$M<0.2~M_\odot$ and $R<1$ km.
If so, PSR 1937+21 can not be an NS or an SS with a crust, but
only a BSS, according to the mass-radius relations of SSs (e.g.,
Xu 2002c).

\begin{acknowledgments}
This work is supported by National Nature Sciences Foundation of
China (10273001) and the Special Funds for Major State Basic
Research Projects of China (G2000077602).
\end{acknowledgments}

\end{document}